\documentclass[12pt]{article}
\usepackage{amsfonts}
\usepackage{latexsym}
\usepackage{amsmath}
\usepackage{amssymb}
\usepackage{amssymb}

\newcommand{\newsection}{    
\setcounter{equation}{0}\section}
\def\appendix#1{\addtocounter{section}{1}\setcounter{equation}{0}
\renewcommand{\thesection}{\Alph{section}}
\section*{Appendix \thesection\protect\indent \parbox[t]{11.15cm}{#1}}
\addcontentsline{toc}{section}{Appendix \thesection\ \ \ #1}}

\newcommand{\be}{\begin{eqnarray}}
\newcommand{\ee}{\end{eqnarray}}
\newcommand{\bea}{\begin{eqnarray}}
\newcommand{\eea}{\end{eqnarray}}
\newcommand{\ba}{\begin{array}}
\newcommand{\ea}{\end{array}}
\newcommand{\nn}{\nonumber \\}

\newenvironment{frcseries}{\fontfamily{frc}\selectfont}{}

\def \la {\label}

\def\cL{{\cal L}}

\def\bbe{{\bf{e}}}

\font\mybb=msbm10 at 11pt

\def\bb#1{\hbox{\mybb#1}}

\def\bR {\bb{R}}

\def\bp {{\bar{p}}}
\def\bq {{\bar{q}}}

\def\hn {{\tilde{\nabla}}}

\long\def\symbolfootnote[#1]#2{\begingroup%
\def\thefootnote{\fnsymbol{footnote}}\footnote[#1]{#2}\endgroup}

\begin{document}
\begin{titlepage}
\begin{center}
\vspace*{-1.0cm}

\vspace{2.0cm} {\Large \bf  IIB black hole horizons with five-form flux and extended
supersymmetry} \\[.2cm]

\vspace{1.5cm}
 {\large  U. Gran$^1$, J. Gutowski$^2$ and  G. Papadopoulos$^{3,}$\symbolfootnote[1]{On study leave from the Department of Mathematics, King's College London, UK.}}

\vspace{0.5cm}

${}^1$ Fundamental Physics\\
Chalmers University of Technology\\
SE-412 96 G\"oteborg, Sweden\\

\vspace{0.5cm}
${}^2$ Department of Mathematics\\
King's College London\\
Strand\\
London WC2R 2LS, UK\\

\vspace{0.5cm}
${}^3$ PH-TH Division\\
 CERN\\
CH-1211 Geneva, Switzerland\\

\vspace{0.5cm}

\end{center}

\vskip 1.5 cm
\begin{abstract}

\end{abstract}

We classify under some assumptions the IIB black hole horizons with 5-form flux preserving more than 2
supersymmetries. We find that the spatial horizon sections with non-vanishing flux preserving 4
supersymmetries are locally isometric  either to $S^1\times S^3\times T^4$ or
to $S^1\times S^3\times K_3$ and the associated near horizon geometries are locally isometric to $AdS_3\times S^3\times T^4$ and
$AdS_3\times S^3\times K_3$, respectively.
The near horizon geometries preserving more than 4 supersymmetries
are locally isometric to $\bR^{1,1}\times T^8$.

\end{titlepage}


\section{Introduction}

There is much evidence that higher dimensional gravitational theories have
black hole solutions with exotic horizon topologies. This is supported by the existence of black rings
in 5-dimensions \cite{ring1}, the results in \cite{gibbons1, rogatko, obers1}, as well as numerous near horizon calculations which have unveiled large classes
of unexpected horizon topologies \cite{hh, kunduri, iibhor}. Interest naturally focuses
on 10- and 11-dimensional supergravities which arise as the
effective theories of strings and M-theory. In particular,  all
near horizon geometries of heterotic supergravity have been found, and those that preserve
half of the spacetime supersymmetry have been classified \cite{hh}. In addition, the geometry
of IIB horizons with 5-form flux preserving at least 2 supersymmetries has been
identified \cite{iibhor}. It is found under certain assumptions that either the near horizon geometry\footnote{It is not apparent that all near horizon geometries can be extended to
 full black hole solutions, see eg \cite{hh, kunduri} for a detailed discussion. In IIB supergravity there are examples of supersymmetric black holes for which their near horizon geometries are those described in \cite{iibhor}.}is a product
${\mathbb{R}}^{1,1}\times X^8$, where $X^8$ is a special holonomy manifold, or the spatial horizon
section ${\cal S}$ is  a  Calabi-Yau manifold with skew-symmetric torsion and the square
of the Hermitian form $\omega$ is $\partial\bar\partial$-closed\footnote{In the terminology of \cite{iibhor}, ${\cal S}$ is a 2-strong Calabi-Yau with torsion manifold or 2-SCYT for short.
The Calabi-Yau condition requires that ${\rm hol}(\hat\nabla)\subseteq SU(4)$ while the 2-strong structure refers to the restriction $\partial\bar\partial\omega^2=0$ on the Hermitian form $\omega$.}, i.e.~$\partial\bar\partial\omega^2=0$. It is remarkable that all
the conditions on ${\cal S}$
which arise from the analysis of the field and  Killing spinor equations (KSEs) of IIB supergravity
  can be described in terms of a connection $\hat\nabla$  with skew-symmetric torsion,   ${\rm hol}(\hat\nabla)\subseteq SU(4)$, even
 though the only active flux is the
5-form.

The presence of a connection with skew-symmetric torsion in IIB horizons with 5-form flux, and so the apparent similarity of their geometries to those which arise in heterotic supergravity,
indicates that there may be a classification of the geometries of all IIB horizons preserving any number of supersymmetries.
This is in analogy with similar  results that have been obtained for the horizons of heterotic supergravity \cite{hh}. However unlike for heterotic supergravity \cite{het}, there is no complete classification of  solutions to the KSEs of IIB supergravity. The solution of the KSEs of IIB supergravity
is known only for backgrounds preserving one supersymmetry \cite{iibn1} and for backgrounds with nearly maximal number of supersymmetries \cite{iib28}. The solution of the KSEs for all IIB horizons with 5-form flux preserving more than 2 supersymmetries, and the corresponding understanding of their geometries, will rely on the special form of the background.

In this paper we shall classify  all IIB near horizon geometries with 5-form flux preserving more than 2 supersymmetries. We shall find that those
preserving 4 supersymmetries with non-vanishing flux are locally isometric to $AdS_3\times S^3\times T^4$ or $AdS_3\times S^3\times K_3$. The associated
spatial horizon sections are $S^1\times S^3\times T^4$ and $S^1\times S^3\times K_3$, respectively. In addition if any near horizon
geometry preserves more than 4 supersymmetries, it is locally isometric to ${\mathbb{R}}^{1,1}\times T^8$.

We have obtained our results under certain assumptions. These assumptions have been explained in detail in section \ref{ass}.
The main role of these assumption is to restrict the choice of  spinors that can appear as Killing spinors for IIB horizons.
In particular, an assumption is used to rule out the presence of a Killing spinor for IIB horizons which lies in
the generic $SU(4)$ class of \cite{iibhor}. This is achieved by either imposing  a certain non-vanishing condition or setting
a component of 5-form flux to vanish. In addition, it is assumed that the Killing vector bi-linear of the Killing spinors
coincides with the stationary Killing vector field of the black hole.

Using these assumptions, we have shown that the Killing spinors of $N=4$ IIB horizons can be chosen to be pure spinors
which have isotropy group $\times^2 SU(2)\ltimes{\mathbb{R}}^8$ in $Spin(9,1)$. In addition, the Killing spinors of $N=6$ IIB horizons
are again pure spinors with isotropy group $U(1)\ltimes {\mathbb{R}}^8$ in $Spin(9,1)$. In fact in both cases, the Killing spinors can be viewed as $Spin(8)$ spinors on the spatial
8-dimensional horizon sections ${\cal S}$ in which case the isotropy groups are $\times^2SU(2)$ and $U(1)$, respectively.
These
kinds of Killing spinors are reminiscent of the Killing spinors that appear in supersymmetric backgrounds of heterotic supergravity, see table 1 of \cite{het1}.
This analogy between Killing spinors in IIB and heterotic supergravities extends to the geometries of supersymmetric
backgrounds. In particular, the spatial horizon sections ${\cal S}$ of IIB horizons preserving 4 supersymmetries admit a hidden
connection $\hat\nabla$ with skew-symmetric torsion such that ${\rm hol}(\hat\nabla)\subseteq \times^2SU(2)$. This extends to the IIB horizons
preserving 6 supersymmetries. However in this case, 8-dimensional manifolds equipped with a connection with skew-symmetric torsion whose holonomy
such that  ${\rm hol}(\hat\nabla)\subseteq U(1)$ have vanishing Riemann curvature \cite{hps}.

Before we proceed with the analysis, it is worth noting that our assumptions  rule out certain near horizon geometries
which are known to exist preserving more than 2 supersymmetries. One example such example is  $AdS_5\times S^5$ which is a maximally
supersymmetric background. This is included in the $N=2$ supersymmetric near horizon geometries of \cite{iibhor}.   But it is
excluded in the classification we give for near horizon geometries with more than 2 supersymmetries.

This paper has been organized as follows. In section 2, the analysis of the IIB KSEs for near horizon geometries with generic
$SU(4)$ invariant  Killing spinors is revisited. In section 3, we explain the assumptions that we use to examine the IIB near horizon
geometries with extended supersymmetry and explore some of their consequences. In section 4, we classify all near horizon geometries
with 5-form flux and describe the similarities with heterotic geometries. In section 5, we show that all IIB near horizon geometries
with 5-form flux preserving more than 4 supersymmetries are locally isometric to ${\mathbb{R}}^{1,1}\times T^8$.

\newsection{N=2 IIB  horizons revisited}

\subsection{Killing spinor equations}

The analysis of the Killing spinor equations has been made in \cite{iibhor}. Here we shall summarize some of the results that will be used in the
rest of the paper. The metric and 5-form field strength of the near horizon geometry written in  Gaussian null co-ordinates are
 \bea
ds^2&=&2 \bbe^+ \bbe^- + \delta_{ij} \bbe^i \bbe^j ~,
\cr
F&=&  r \bbe^+ \wedge (dY-h \wedge Y)
+ \bbe^+ \wedge \bbe^- \wedge Y + \star_8 Y~,
\label{fivef}
\eea
where
\be
\label{basis1}
\bbe^+ = du, \qquad \bbe^- = dr + rh -{1 \over 2} r^2 \Delta du, \qquad \bbe^i = e^i_I dy^I~,
\ee
and the $r,u$-dependence of the components is explicitly stated. Therefore $\Delta$, $h$ and $Y$ depend only on the coordinates, $y$, of the spatial horizon section, ${\cal S}$.  ${\cal S}$
is the co-dimension 2 submanifold defined by $r=u=0$ and it is assumed to be closed, i.e.~compact without boundary. For more explanation about our conventions see \cite{iibhor}.

The KSE equation of IIB supergravity \cite{west, schwarz, howe} with only 5-form flux is
\be
\label{kse}
\nabla_M\epsilon
+{i \over 48} F_{M N_1 N_2 N_3 N_4}\Gamma^{N_1 N_2 N_3 N_4}  \epsilon =0~,
\ee
where $\nabla$ is the spin connection associated with the frame (\ref{basis1}) and $\epsilon$ is a spinor in the positive chirality
complex Weyl representation of $Spin(9,1)$. To solve the KSE, we first identify the dual 1-forms
\be
V = -{1 \over 2} r^2 \Delta \bbe^+ + \bbe^- \ ,
\ee
and
\be
Z = \langle B (C \epsilon^*)^*, \Gamma_A \epsilon \rangle
= \langle \Gamma_0 \epsilon, \Gamma_M \epsilon \rangle~\bbe^A~.
\la{kill}
\ee
of the two Killing vector fields. The first vector field  is the stationary Killing vector field  $\partial_u$  of the black hole,
 and the other is the Killing vector field constructed as a Killing spinor bi-linear. In such case, the KSE can be solved along the light-cone directions
 to find that the Killing spinor can be expressed as
 \bea
 \epsilon=\eta_++r \Gamma_- \bigg({1 \over 4} h_i \Gamma^i
+{i \over 12} Y_{n_1 n_2 n_3} \Gamma^{n_1 n_2 n_3} \bigg) \eta_+~,~~~\Gamma_+\eta_+=0~,
\eea
 where $\eta_+$ is an even-chirality $Spin(8)$ spinor which depends only on the coordinates of ${\cal S}$.

Up to  $Spin(8)$ $r,u$-independent gauge transformations \cite{iibn1},
one can take without loss of generality\footnote{For our spinor conventions as well as for the definition  of form spinor bi-linears see \cite{iibn1}.}
\be
\eta_+=p+q e_{1234}~,
\la{pqe}
\ee
where $p,q$ are complex functions of ${\cal S}$. In such a case, one finds that
$|p|^2+|q|^2$ must be a (non-zero) constant and
\be
\label{hexp}
h_i =- {|p|^2-|q|^2 \over |p|^2+|q|^2} Y_{i \ell_1 \ell_2} \omega^{\ell_1 \ell_2}~,
\ee
where the Hermitian form $\omega$ on ${\cal{S}}$ is
\be
\omega = -\bbe^1\wedge \bbe^6 - \bbe^2\wedge \bbe^7 - \bbe^3\wedge \bbe^8-\bbe^4\wedge \bbe^9~.
\ee
Moreover,
\be
\label{dexp3}
\Delta = {2 \over 3} {\hat{Y}}_{\ell_1 \ell_2 \ell_3} {\hat{Y}}^{\ell_1 \ell_2 \ell_3}~,
\ee
where
\be
\label{dexp2}
{\hat{Y}}_{\ell_1 \ell_2 \ell_3}= (Y_{(0,3)}+Y_{(3,0)})_{\ell_1 \ell_2 \ell_3}
-{i \over 8 (|p|^2+|q|^2)} Y_{m n_1 n_2} \omega^{n_1 n_2}
\bigg( p \bar{q} \chi^m{}_{\ell_1 \ell_2 \ell_3} - {\bar{p}} q {\bar{\chi}}^m{}_{\ell_1 \ell_2 \ell_3} \bigg),
\ee
and
\be
\chi = (\bbe^1+i \bbe^6)\wedge (\bbe^2+i\bbe^7) \wedge (\bbe^3+i \bbe^8)\wedge (\bbe^4+i \bbe^9)~,
\ee
is the $(4,0)$ form on ${\cal{S}}$. So $\Delta \geq 0$, as expected.

Furthermore  the remaining components of the KSE imply that
\be
\label{kse2}
\hn_i \eta_+ -{1 \over 4} h_i \eta_+ -{i \over 12} Y_{\ell_1 \ell_2 \ell_3}\Gamma^{\ell_1 \ell_2 \ell_3} \Gamma_i \eta_+=0~,
\ee
and
\bea
\label{alg6a}
\bigg( \big[{1 \over 4} \hn_j h_i -{1 \over 8} h_i  h_j +{1 \over 4} Y_{i q_1 q_2} Y_j{}^{q_1 q_2} \big] \Gamma^j
+ \big[{i \over 12}(\hn_i Y_{\ell_1 \ell_2 \ell_3}- (dY)_{i \ell_1 \ell_2 \ell_3})
\nn
+{i \over 24} \big( (h \wedge Y)+ \star_8 (h \wedge Y) \big)_{i \ell_1 \ell_2 \ell_3}
-{1 \over 144} Y_{i m_1 m_2} Y_{m_3 m_4 m_5} \epsilon^{m_1 m_2 m_3 m_4 m_5}{}_{\ell_1 \ell_2 \ell_3}
\nn
-{1 \over 4} Y_{m [\ell_1 \ell_2} Y_{\ell_3] i}{}^m \big] \Gamma^{\ell_1 \ell_2 \ell_3} \bigg) \eta_+=0~,
\nn
\eea
where $\hn$ denotes the Levi-Civita connection on ${\cal{S}}$.

Also, on expanding out ({\ref{kse2}}),
one obtains the conditions:
\bea
\label{kse2a}
\partial_\alpha p + \big({1 \over 2} \Omega_{\alpha, \beta}{}^\beta - iY_{\alpha \beta}{}^\beta -{1 \over 4} h_\alpha
\big)p &=&0
\nn
\partial_\alpha \bp + \big(-{1 \over 2} \Omega_{\alpha, \beta}{}^\beta -{1 \over 4} h_\alpha \big)\bp
-{i \over 3} \epsilon_{\alpha \gamma_1 \gamma_2 \gamma_3} Y^{\gamma_1 \gamma_2 \gamma_3} \bq &=&0
\nn
\partial_\alpha q + \big(-{1 \over 2} \Omega_{\alpha, \beta}{}^\beta -{1 \over 4} h_\alpha \big)q +{i \over 3}
\epsilon_{\alpha \gamma_1 \gamma_2 \gamma_3} Y^{\gamma_1 \gamma_2 \gamma_3} p &=&0
\nn
\partial_\alpha \bq + \big({1 \over 2} \Omega_{\alpha, \beta}{}^\beta + iY_{\alpha \beta}{}^\beta -{1 \over 4} h_\alpha
\big) \bq &=&0
\eea
and
\bea
\label{kse2b}
\Omega_{\alpha, \gamma_1 \gamma_2} \epsilon^{\gamma_1 \gamma_2}{}_{\bar{\delta}_1 \bar{\delta}_2}
 &=& {4 p \bq \over |p|^2+|q|^2} \Omega_{\alpha, \bar{\delta}_1 \bar{\delta}_2}
\nn
i Y_{\alpha \bar{\delta}_1 \bar{\delta}_2} -i \delta_{\alpha [\bar{\delta}_1} Y_{\bar{\delta}_2] \beta}{}^\beta
&=& {(|p|^2-|q|^2) \over 2 (|p|^2+|q|^2)} \Omega_{\alpha, \bar{\delta}_1 \bar{\delta}_2} \ .
\eea

There are three special cases to consider which are distinguished by the choice  $\eta_+$ which in turn put
restrictions on the functions $p$ and $q$. In what follows we shall focus on the generic $SU(4)$ case for which the spinor
$\eta_+$ is chosen as
\bea
\eta_+=p \, 1+q e_{1234}, ~~~p \neq 0~,~~~q \neq 0~,~~~|p|^2-|q|^2 \neq 0~.
\eea
The remaining two cases have been exhaustively examined in \cite{iibhor}.

\subsection{Generic $SU(4)$ invariant Killing spinors revisited}

For solutions for which $\eta_+$ is a generic $SU(4)$ invariant Killing spinor, we proceed by
considering the +- component of the Einstein equation \cite{iibhor}.
This equation can be rewritten, on using ({\ref{dexp3}}), as
\bea
\label{einaux}
\hn^i h_i &=& -2 {\big(|p|^2 |q|^2-{1 \over 2} \big) \over \big(|p|^2-|q|^2 \big)^2 } h^2 -8
Y_{{\bar{\delta}} \sigma_1 \sigma_2} Y^{{\bar{\delta}} \sigma_1 \sigma_2}
\nn
&+&{4i \over 3 (|p|^2-|q|^2)} \bigg( p \bar{q} \epsilon^{{\bar{\delta}}_1
{\bar{\delta}}_2 {\bar{\delta}}_3 {\bar{\delta}}_4} h_{{\bar{\delta}}_1}
Y_{{\bar{\delta}}_2 {\bar{\delta}}_3 {\bar{\delta}}_4}
- {\bar{p}} q \epsilon^{\delta_1 \delta_2 \delta_3 \delta_4} h_{\delta_1} Y_{\delta_2 \delta_3 \delta_4} \bigg) \ .
\eea

We make use of the following identities obtained from ({\ref{kse2a}}) and ({\ref{kse2b}}):
\bea
h_\alpha = (|q|^2-|p|^2) \partial_\alpha \log \big( {p \over \bar{q}} \big),
\qquad \Omega_{\bar{\gamma},}{}^{\bar{\gamma}}{}_\alpha = {1 \over 2(|q|^2-|p|^2)}
\partial_\alpha \log \big( {p \over \bar{q}} \big) \ ,
\eea
and
\bea
Y_{\delta_1 \delta_2 \delta_3} = -{i \over 2} p {\bar{q}} \epsilon_{\delta_1 \delta_2 \delta_3}{}^{\bar{\alpha}}
\partial_{\bar{\alpha}} \log \big( {p \over \bar{q}} \big) \ ,
\eea
and set
\be
p = |p| e^{i \phi}, \qquad q=|q|e^{i \psi} \ ,
\ee
for real $\phi, \psi$.
On substituting these expressions back into ({\ref{einaux}}), one finds, after some manipulation, that

\bea
\label{intv3}
{\tilde{\nabla}}^2 \bigg( \big(|p|^2 |q|^2 \big)^{-{1 \over 4}} \bigg)
&=& \big(|p|^2 |q|^2 \big)^{-{1 \over 4}} \bigg(
{1 \over 4}(|p|^2-|q|^2)^2 {\tilde{\nabla}}^i (\phi+ \psi)
{\tilde{\nabla}}_i (\phi+ \psi)
\nn
&+&{h^2 \over 12 (|p|^2-|q|^2)^2}
+4 {\tilde{Y}}_{\bar{\delta} \sigma_1 \sigma_2} {\tilde{Y}}^{\bar{\delta} \sigma_1 \sigma_2}
\nn
&+&{1 \over 2 |p|^2|q|^2 (|p|^2-|q|^2)^2} \hn_i (|p|^2|q|^2) \hn^i (|p|^2|q|^2) \bigg)~,
\eea
where ${\tilde{Y}}$ denotes the traceless part of the $(1,2)+(2,1)$ of $Y$.
Note also that we have extracted the trace terms from the term quadratic in $Y$ in
({\ref{einaux}}) and rewritten their contribution in terms of $h^2$.

To explore the consequences of (\ref{intv3}), one needs that $ \big(|p|^2 |q|^2 \big)^{-{1 \over 4}}$ is a smooth function on ${\cal S}$. For this
$p$ and $q$ must be smooth no-where vanishing functions on ${\cal S}$.   The spatial horizon section ${\cal S}$
admits an $SU(4)$ structure. So having chosen a trivialization using the globally defined sections $1$ and $e_{1234}$
of the spinor bundle, $p$ and $q$ can be chosen as globally defined smooth functions on ${\cal S}$. Moreover, the parallel transport
equation (\ref{kse2a}) implies that   $|p|^2+|q|^2$ is constant. So
 although $p$ and $q$ cannot simultaneously vanish  as  $|p|^2+|q|^2\not=0$, in general the parallel
 transport equation (\ref{kse2a}) allows for $p$ or $q$ to  have a vanishing locus on ${\cal S}$. To exclude this possibility, one
 has to make an additional assumption. For this, one can simply assume that $\eta_+$ is a no-where pure spinor on ${\cal S}$. Alternatively, one
 can use the parallel transport equation (\ref{kse2a}) and set the (3,0) part of $Y$ to zero, $Y^{3,0}=0$. In such case, the parallel transport equation
 (\ref{kse2a}) factorizes to one for $p$ and another one for $q$. So if $p$ or $q$  vanish at one point, then they vanish everywhere on ${\cal S}$.
 Thus if $Y^{3,0}=0$ and $\eta_+$ is generic, then it is no-where pure.

Assuming that $ \big(|p|^2 |q|^2 \big)^{-{1 \over 4}}$ is smooth, noting that the RHS of ({\ref{intv3}}) is
 non-negative  and  using the maximum principle on
({\ref{intv3}}),  one finds that $|p|$,  $|q|$
and
$\psi+\phi$ are constant. Furthermore $h=0$ and ${\tilde{Y}}=0$, and
so the $(2,1)$ part of $Y$ vanishes. On substituting all of these conditions back into
({\ref{kse2a}}), one also finds that the $(3,0)$ part of $Y$ vanishes as well.  So $Y=0$,
and hence $\Delta=0$.

Thus we have shown that if $\eta_+=p1+q e_{1234}$ is a generic $SU(4)$ invariant no-where pure spinor,
then  $p$, $q$ must be constant and the flux  $F$ vanishes. The spacetime is a product
$\bR^{1,1} \times {\cal{S}}$, where ${\cal{S}}$ is a compact Calabi-Yau 4-fold.

\newsection{Horizons with more than 2 supersymmetries}

\subsection{Additional Killing spinors}
\label{ass}

The solutions of the KSEs of IIB supergravity for backgrounds with more than 2 supersymmetries have not been classified, other than in cases for which the amount of supersymmetry preserved is near-maximal.
Nevertheless, we can solve the KSEs
of IIB supergravity for near horizon geometries preserving more than 2 supersymmetries
by relying on the special form of the backgrounds  and on some additional assumptions
that we shall make. In particular, we take that
\begin{itemize}

\item[(i)] the 1-form bilinears of all the Killing spinors, and so their linear combinations,
are proportional to the 1-form whose dual vector field is  ${\partial \over \partial u}$,

\item[(ii)] and all of the Killing spinors and their linear combinations are constructed from pure spinors $\eta_+$.

\end{itemize}

The first assumption is needed in order for the analysis we have done for one linearly independent Killing spinor in \cite{iibhor} to apply
for all additional Killing spinors. As has been explained, the starting point of the analysis of the KSE
of backgrounds with 2 supersymmetries is the identification of the Killing spinor 1-form bilinear with the 1-form dual to the stationary
Killing vector field of the black hole solution $\partial_u$.

The second assumption is motivated by the results of
the previous section. Any additional Killing spinor $\epsilon$ must be  associated with either a  $Spin(7)$ invariant, or a generic $SU(4)$ invariant, or
a pure $SU(4)$ invariant spinor $\eta_+$. In the first case, the horizon is a product with vanishing 5-form flux. The same is
true for the second case provided the assumptions we have made in the previous section are valid. Thus the only possibility
that can arise yielding non-product horizons is that for which the Killing spinor is constructed from a pure $SU(4)$ invariant
spinor $\eta_+$. Hence assumption (ii) follows from assumption (i) and the hypothesis used in the previous section
to understand the near horizons geometries associated with  generic $SU(4)$ invariant spinors $\eta_+$.

Utilizing both (i) and (ii), any additional Killing spinor $\epsilon$ of a near horizon geometry is $r$ and $u$ independent. In particular,  $\epsilon=\eta_+$, and so
from the results of \cite{iibhor}, it satisfies
\be
\label{eq2}
\big( h_i \Gamma^i +{i \over 3} Y_{\ell_1 \ell_2 \ell_3} \Gamma^{\ell_1 \ell_2 \ell_3} \big) \epsilon =0~,
\ee
\be
\label{eq3}
dh_{ij} \Gamma^{ij} \epsilon =0~,
\ee
\be
\label{eq4}
\hn_i \epsilon  +\big(-{1 \over 2} h_i -{i \over 2} Y_{i mn} \Gamma^{mn} -{1 \over 4} h_j \Gamma_i{}^j \big)
\epsilon =0~.
\ee
Note that $\Delta=0$.

\subsection{N=2 Solutions}
\label{twosol}

We briefly summarize  the N=2 solutions \cite{iibhor}, working in a holomorphic basis on ${\cal{S}}$ which will
be convenient for subsequent analysis.
The $N=2$ solutions have Killing spinors which can be taken, without loss of generality to be
\be
\epsilon^1= 1, \qquad \epsilon^2=i1
\ee
The conditions obtained on $h$, $Y$ and the spin connection are as follows:
\be
dh_{\alpha \beta}=0, \qquad dh_\alpha{}^\alpha =0
\la{con1}
\ee
\be
Y_{\alpha_1 \alpha_2 \alpha_3}=0, \qquad \Omega_{\alpha, \beta}{}^\beta -iY_{\alpha \beta}{}^\beta =0,
\qquad i Y_{\alpha \beta}{}^\beta + {1 \over 2} h_\alpha =0~,
\la{con2}
\ee
\be
\Omega_{\alpha, \beta_1 \beta_2}=0~,~~~\Omega_{\bar\beta,}{}^{\bar\beta}{}_\alpha+ \Omega_{\alpha, \beta}{}^\beta=0~,
\la{con3}
\ee
and
\be
Y = {1 \over 4} \big( d \omega -h \wedge \omega\big)~,
\la{con4}
\ee
where
\be
\omega= -i \big( \bbe^1 \wedge \bbe^{\bar{1}}+ \bbe^2 \wedge \bbe^{\bar{2}}+ \bbe^3 \wedge \bbe^{\bar{3}}
+ \bbe^4 \wedge \bbe^{\bar{4}} \big)~.
\la{her1}
\ee
As  has been explained in \cite{iibhor} the spatial horizon section ${\cal S}$ has a hidden Calabi-Yau with torsion structure.

\newsection{N=4 horizons}

\subsection{Additional Killing spinors}

The first two Killing spinors of $N=4$ solutions are identified with those of $N=2$ backgrounds. As we have demonstrated a basis
for these can be chosen as $\epsilon^1=1, \epsilon^2=i \epsilon^1$. As the IIB KSEs with only 5-form flux are linear over the complex numbers,
if the third Killing spinor is $\epsilon^3$, then the fourth can be chosen as $\epsilon^4=i\epsilon^3$.
 Therefore in order to identify the  two additional Killing spinors, it  suffices to choose the third. For this first observe that
  because of the assumptions we have made in the previous section
  any additional Killing spinor must be pure. As a consequence a Killing spinor $\epsilon$ is identified with the associated spinor
  $\eta_+$ on the spatial horizon section ${\cal S}$. Thus the third Killing spinor can be chosen up to gauge transformations
  of the spatial horizon section. As the first two Killing spinors have isotropy group $SU(4)$, the third Killing spinor can be chosen
  up to $SU(4)$ gauge transformations. In particular, an analysis of the orbits of $SU(4)$ on the positive chirality $Spin(8)$ spinors reveals
  that
\be
\epsilon^3= p \,1 +q\, e_{1234}+ \alpha\, e_{12}+\beta\, e_{34}~,
\ee
where $p,q, \alpha$ and $\beta$ are complex functions on ${\cal S}$.

The spinor $\epsilon^3$ can be simplified using the assumptions of section \ref{ass}. In particular
we require that $\epsilon^3+ \lambda\, \epsilon^1$ must be a pure spinor for any choice of
constant complex parameter $\lambda$.  This restriction can be imposed by setting
the 1-form\footnote{Since the spinors are complex, one can construct three independent 1-form spinor bilinears. One of them is associated
with the Killing vector of supersymmetric IIB backgrounds and the other two vanish for pure spinors. Compare (\ref{vas}) with the bilinear associated with
the Killing vector (\ref{kill}).}  spinor bilinear
\be
\langle B (\epsilon^3+\lambda\, \epsilon^1)^*, \Gamma_A (\epsilon^3+\lambda\, \epsilon^1) \rangle \, \bbe^A~,
\la{vas}
\ee
to vanish for all $\lambda$.
This implies that
\be
q=0,  \qquad \alpha \beta=0~.
\ee
However, it is furthermore straightforward to show that $p\,1+\alpha\, e_{12}$ and $p\,1 + \beta\, e_{34}$
are in the same orbit of $SU(4)$. Hence, without loss of generality, we take
\be
\epsilon^3 = p\,1 + \alpha\, e_{12}~.
\ee
Some further simplification is possible. For this,  we compute the 1-form  spinor bilinear
\be
\kappa  = \langle B (C*(\epsilon^3+\lambda\,\epsilon^1))^*, \Gamma_A (\epsilon^3+\lambda, \epsilon^1) \rangle\, \bbe^A=-\sqrt{2}(|p+\lambda|^2+|\alpha|^2) \bbe^-~,
\ee
which is  dual to a Killing vector field for any choice of constant parameter $\lambda$. The Killing vector equations imply that
 $|p+\lambda|^2+|\alpha|^2$ is independent of the coordinates of ${\cal S}$ and so constant for any $\lambda$. This in turn implies that
$p$ and $\alpha$ are constant. Hence up to a $SU(4)$ gauge transformation, a basis of the four Killing spinors can be chosen as
\be
\epsilon^1=1~,~~~\epsilon^2=i\, 1~,~~~~\epsilon^3= e_{12}~,~~~~\epsilon^4=i\, e_{12}~.
\ee
The isotropy group of the 4 Killing spinors in $Spin(9,1)$ is $\times^2SU(2)\ltimes \bR^8$ while in $Spin(8)$ is $\times^2SU(2)$.

\subsection{Solution of KSEs}

To continue with the analysis, we have to solve the KSEs for the spinor $\epsilon^3= e_{12}$. This can be done directly as for the $N=2$ backgrounds. However because of the simplicity
of $\epsilon^3$, it is possible to just read off the restrictions imposed by the KSE on the fields from those of backgrounds with two supersymmetries.
For this observe that all  restrictions that arise from the KSE in $N=2$ backgrounds can be expressed as conditions on the fields written in a holomorphic basis
with respect to a Hermitian form $\omega$ which is computed from the 3-form Killing spinor bi-linear. These are summarized in section \ref{twosol}.

To find the conditions on the fields imposed by the $e_{12}$ Killing spinor, we calculate  the 3-form bi-linear of $e_{12}$ to find
\be
\Theta = \sqrt{2} i \bbe^- \wedge \omega'~,
\ee
where
\be
\omega' = i \big( \bbe^1 \wedge \bbe^{\bar{1}}+ \bbe^2 \wedge \bbe^{\bar{2}}- \bbe^3 \wedge \bbe^{\bar{3}}
- \bbe^4 \wedge \bbe^{\bar{4}} \big)~.
\ee
It is clear that the Hermitian forms $\omega$ and $\omega'$ are related by a rotation which exchanges the first two holomorphic directions with the corresponding
anti-holomorphic ones. The associated complex structures commute.  Therefore the conditions imposed by the Killing spinor $\epsilon^3$ are as those in section \ref{twosol} for the spinor $\epsilon^1$
 but now with the first two holomorphic
 indices replaced by anti-holomorphic ones and vice versa. To implement this, we  split the holomorphic indices as $\alpha = (a, \mu )$
for $a=1,2$, $\mu=3,4$, however, middle Roman  indices $i,j$ and the first Greek indices $\alpha, \beta, \gamma$ are reserved for real indices  and all the
holomorphic indices on ${\cal{S}}$, respectively.  The conditions that arise from the KSE imposed on $\epsilon^1$ are those summarized in section \ref{twosol} by simply setting  $\alpha=(a, \mu )$
and similarly for the rest of the indices. While the conditions that arise from the KSE imposed on $\epsilon^3$ are again given by those in \ref{twosol} by now for
$\alpha=(\bar a, \mu)$ and similarly for the rest of the indices.

\subsection{Analysis of the conditions}

 Using the procedure explained in the previous section for solving the KSE for both spinors $\epsilon^1$ and $\epsilon^3$, we find that condition that the (3,0) part of $Y$ vanishes in (\ref{con2}) with respect to both complex structures implies that the only non-vanishing
components of $Y$ up to complex conjugation are
\bea
Y_{a {\bar{b}}_1 {\bar{b}}_2}~,~~~ Y_{a {\bar{b}} {\bar{\mu}}}~, ~~~Y_{\bar{a} \mu {\bar{\nu}}}~,~~~
Y_{\mu {\bar{\nu}}_1 {\bar{\nu}}_2}~.
\eea
In particular, $Y_{{a} \bar\mu {\bar{\nu}}}=0$. In addition the last condition in (\ref{con2}) implies that
\bea
Y_{i a}{}^a =0, \qquad h_i = -2i Y_{i\mu}{}^\mu~.
\eea

To proceed further, it will be convenient to set
\be
\omega_1 =  -i \big( \bbe^1 \wedge \bbe^{\bar{1}}+ \bbe^2 \wedge \bbe^{\bar{2}} \big),
\quad
\omega_2 =  -i \big( \bbe^3 \wedge \bbe^{\bar{3}}+ \bbe^4 \wedge \bbe^{\bar{4}} \big)~,
\ee
and decompose
\be
h = h_1 + h_2~.
\ee
where $(h_1)_\mu=0$, $(h_2)_a=0$.
We also write
\be
Y = {\mathring {Y}} - \phi \wedge \omega_2~,~~~\phi = {1 \over 4} h_1 + {1 \over 2} h_2~,
\ee
such that ${\mathring{Y}}$ is traceless w.r.t both $\omega_1$ and $\omega_2$.

Next, consider the $+-$ component of the Einstein equations \cite{iibhor}, which for $\Delta=0$, is
\be
\hn^i h_i = h^2 - {4 \over 3} Y_{i_1 i_2 i_3} Y^{i_1 i_2 i_3}~.
\ee
This can be rewritten as
\be
\hn^i h_i = -(h_2)^2 -{4 \over 3} {\mathring{Y}}_{i_1 i_2 i_3} {\mathring{Y}}^{i_1 i_2 i_3}~.
\ee
Integrating over ${\cal S}$ and using that ${\cal S}$ is compact without boundary, one finds that
\be
\hn^i h_i =0~,
\ee
and
\be
h_2 =0,  \qquad {\mathring{Y}}=0~.
\ee
Using these results, $Y$ can be expressed as
\be
\label{ysol}
Y= -{1 \over 4} h \wedge \omega_2~,~~~h=h_1~.
\la{newy}
\ee

Furthermore $Y$ can be expressed both in terms of $\omega$ and $\omega'$ (\ref{con4}). Writing the two expressions in terms of $\omega_1$ and $\omega_2$ and comparing them with
(\ref{newy}), one finds that
\be
\label{cv1}
d \omega_2 =0,  \qquad d \omega_1 - h \wedge \omega_1 =0~.
\ee
Combining, the first condition in (\ref{con3}) for both complex structures with the first condition in (\ref{cv1}), one finds that the non-vanishing
components of the frame connection $\Omega$ up to a complex conjugation are
\bea
\Omega_{a,b\bar c}~,~~~\Omega_{a,\bar b\bar c}~,~~~\Omega_{a,\mu\bar \nu}~,~~~\Omega_{\mu, a \bar b}~,~~~\Omega_{\mu,\nu\bar\lambda}~.
\la{Ocom}
\eea
These are in addition restricted by the second equation in (\ref{con3}) for both complex structures. In particular, one finds that
\bea
\Omega_{i,\nu}{}^\nu=\Omega_{\mu, a}{}^a=0~,~~~\Omega_{\nu,\mu}{}^\nu=\Omega_{\bar a, \mu}{}^{\bar a}=\Omega_{a, \mu}{}^a=0~,~~~ \Omega_{a,b}{}^b-\Omega_{\bar b,a}{}^{\bar b}=0~.
\eea
It is clear from these that
\be
d(\bbe^3 \wedge \bbe^4)=0~,
\ee
and also
\be
\label{cv2}
d (\bbe^1 \wedge \bbe^2) - h \wedge (\bbe^1 \wedge \bbe^2) =0~.
\ee
The integrability conditions of ({\ref{cv1}}) and ({\ref{cv2}}) imply that
\be
dh \wedge \omega_1 =0, \qquad dh \wedge \bbe^1 \wedge \bbe^2=0~,
\ee
and so
\be
dh = dh_{a \bar{b}}\, \bbe^a \wedge \bbe^{\bar{b}}, \qquad dh_a{}^a=0~.
\ee

The rotation is further restricted. In fact, $h$ is a parallel 1-form on ${\cal S}$. For this note that
\be
\star dh = {1 \over 2}  dh \wedge \omega_2 \wedge \omega_2~.
\ee
Using this, one has that
\be
I={1\over2} \int_{\cal{S}} dh_{ij} dh^{ij} =\int_{\cal{S}}\, dh \wedge \star dh={1\over2} \int_{\cal{S}} dh \wedge dh \wedge \omega_2 \wedge \omega_2=0~,
\ee
where in the last step we have used that $d\omega_2=0$. Hence
\be
dh=0~,
\ee
and so $h$ is closed. Since $h$ is also co-closed, it implies that $h$ is harmonic.

To proceed, consider the $i,j$ component of the Einstein equations \cite{iibhor}, which can be expressed as
\be
{\tilde{R}}_{ij} = - \hn_{(i} h_{j)} +{1 \over 2} h_i h_j -4 Y_{i n_1 n_2} Y_j{}^{n_1 n_2}
+{2 \over 3} \delta_{ij} Y_{n_1 n_2 n_3} Y^{n_1 n_2 n_3}
\ee
and define
\be
\tilde I = \int_{\cal{S}} \hn_{(i} h_{j)} \hn^{(i} h^{j)}~.
\ee
On integrating by parts, and using $dh=0$, $\hn^i h_i=0$, one finds that
\be
\tilde I = -{1\over2} \int_{\cal S} \big (h_i\tilde \nabla^2 h^i+ \tilde R_{ij} h^i h^j\big)= - \int_{\cal{S}} {\tilde{R}}_{ij} h^i h^j~.
\ee
Using the Einstein equation to express the Ricci tensor in terms of the fluxes and the expression of $Y$ in terms of $h$ (\ref{newy}), we have
\bea
{\tilde{R}}_{ij} h^i h^j=-{1\over 2} \tilde \nabla_i(h^2 h^i)~.
\eea
Thus $\tilde I=0$ and so $h$ is Killing. But also $dh=0$, and so $h$ is parallel
\bea
\tilde \nabla h=0~.
\eea

To proceed, we decompose the metric of spatial horizon section as
\bea
ds^2=ds_1^2+ds^2_2~,~~~ ds^2_1=2 \bbe^1 \bbe^{\bar 1}+2\bbe^2 \bbe^{\bar 2}~,~~~ds^2_2=2 \bbe^3 \bbe^{\bar 3}+2\bbe^4 \bbe^{\bar 4}~.
\eea
This is a product decomposition, i.e.~${\cal S}$ locally metrically decomposes into a product of two hyper-Hermitian 4-dimensional  manifolds ${\cal S}= X \times Y$, where
$X$ is equipped with metric $ds_1^2$, Hermitian  $\omega_1$ and (2,0) $\bbe^1\wedge \bbe^2$ forms, and similarly for $Y$.
This can be easily seen from (\ref{Ocom}). In particular the metric $ds^2_1$, as well as the associated Hermitian forms, are invariant under the action spanned by the dual vector fields
spanned by the frames  $\bbe^3$ and $\bbe^4$ and their conjugates, and similarly for the metric $ds^2_2$ and its associated Hermitian forms.

Furthermore $Y$ is hyper-K\"ahler. This can be seen from $d\omega_2=d(\bbe^3\wedge \bbe^4)=0$ or alternatively from $d\omega_2=0$ and by showing that it is Ricci flat.
The latter follows from the Einstein field equations. Therefore $Y$ is locally isometric to either $T^4$ or $K_3$.

To identify $X$, we first observe that $h$ is a parallel 1-form on $X$. As a result, if $h \neq 0$, 
the metric decomposes as
\bea
ds^2_1=k^{-2} h\otimes h+ds^2(\Sigma)~,
\eea
 where $k^2$ is the constant square length of $h$. It remains to identify the 3-dimensional manifold  $\Sigma$. For this, we evaluate the Ricci tensor along the directions in $X$
 perpendicular to $h$ to find
 \bea
 \tilde R_{ij} V^i W^j={1\over 2} k^2 \delta_{ij} V^i W^j~,~~~ h_i V^i= h_i W^j=0~.
 \eea
Thus $\Sigma$ has constant positive curvature and so it is locally
isometric to $S^3$. It is well known that $S^1\times S^3$ admits an
HKT and so hyper-Hermitian structure. Thus the spatial horizon
section is locally isometric to the product ${\cal S}= S^1\times
S^3\times Y$, where $Y=T^4$ or $K_3$. In turn the near horizon
geometry is isometric to either $AdS_3\times S^3\times T^4$ or
$AdS_3\times S^3\times K_3$.
Adapting local coordinates as
$h=d\phi$, the  full spacetime metric is
\be
ds^2 = 2 du (dr+ k^2 r
d \phi) + k^2 d \phi^2+ k^{-2}\big((\sigma^1)^2+(\sigma^2)^2+(\sigma^3)^2\big)+
ds_2^2~,
\ee
where $\sigma^r$, $r=1,2,3$ is the left-invariant frame
on $S^3$, i.e.~$d\sigma^1=\sigma^2\wedge \sigma^3$ and cyclicly in $1,2$ and $3$. Observe that the radii of $AdS_3$ and $S^3$ are equal.

\subsection{Hidden torsion}

We have demonstrated in \cite{iibhor} that the spatial horizon sections of near horizon geometries which preserve two supersymmetries admit a 2-SCYT structure.
This structure describes the full set of  conditions imposed by the KSEs and the field equations of the supergravity theory on the spatial horizon sections.
To see whether this is the case for spatial horizon sections admitting four supersymmetries consider 8-dimensional manifolds $M$ equipped with a connection
 $\hat\nabla$ with skew-symmetric torsion
$H$ such that ${\rm hol}(\hat\nabla)\subseteq \times^2SU(2)$. The holonomy has been chosen to be $\times^2SU(2)$ because it is the isotropy
group of the Killing spinors in $Spin(8)$. Such manifolds admit two commuting almost complex structures $I$ and $I'$ which we shall assume are simultaneously
integrable. In such a case, it has been shown in \cite{hps} that $M$ locally metrically decomposes as $M=X\times Y$, where $X$ and $Y$ are 4-dimensional KT manifolds.
Since the holonomy is in  $\times^2SU(2)$, it turns out that $X$ and $Y$ are HKT manifolds. The skew-symmetric torsion $H$ is given by
\bea
H=-i_Id\omega~,
\la{skewt}
\eea
where $\omega$ is the Hermitian form of $I$.

It is now clear that the spatial horizon sections ${\cal S}$ admit a connection with skew-symmetric torsion $\hat\nabla$ and ${\rm hol}(\hat\nabla)\subseteq \times^2SU(2)$.
The shew-symmetric torsion is given in (\ref{skewt}).
However, the geometry of ${\cal S}$ is further restricted because of the field equations. These force $Y$ to be hyper-K\"ahler and $X$ to be locally isometric
to $S^1\times S^3$ equipped with the HKT structure.

The existence of hidden skew-symmetric torsion compatible with the geometric data of spatial horizon sections preserving 4 supersymmetries leads
to a prediction. If there is a hidden skew-symmetric torsion structure for spatial horizon sections preserving more than 4 supersymmetries, then
 the near horizon geometry is flat. This conclusion can be reached from the results of \cite{hps}. In particular,  it has been proven that complex manifolds equipped
 with a connection with skew-symmetric torsion whose holonomy is a suitable subgroup of $\times^2 SU(2)$  are flat. The subgroup in $\times^2 SU(2)$
 is an isotropy group of Killing spinors. We shall see that this is indeed the case. Provided that the assumptions
 of section \ref{ass} hold, all near horizon geometries preserving more than 4 supersymmetries are flat.

\newsection{$N \geq 6$ horizons}

\subsection{Killing spinors}
To begin, consider near horizon geometries preserving 6 supersymmetries. Provided that the assumptions of section \ref{ass} hold, the  first 4 Killing spinors
can be chosen as those for the $N=4$ solutions, i.e.~a basis is  $\{1, i\, 1, e_{12}, i e_{12} \}$. Because of the linearity of the KSEs over the complex numbers, if the 5th Killing
   spinor is $\epsilon^5$, then the 6th can be chosen as $i \epsilon^5$. Using again the assumptions in section \ref{ass}, $\epsilon^5$ is a $Spin(8)$ even chirality
   pure spinor on ${\cal S}$. Thus it can be written as
\be
\epsilon^5 = p\, 1 +q\, e_{1234} +{1 \over 2} z^{\alpha \beta} e_{\alpha \beta}~,
\ee
where $p,q, z$ are complex functions on ${\cal S}$. Since any linear combination of the Killing spinors must be pure, we first
 require that $\epsilon^5+\lambda\,1+\mu\, e_{12}$ should be pure, for all possible choices of constant complex parameters
 $\lambda, \mu$.  This implies that the 1-form spinor bi-linear
\be
\label{pure2}
\langle B (\epsilon^5+\lambda\,1+\mu\, e_{12})^*, \Gamma_A (\epsilon^5+\lambda\,1+\mu\, e_{12}) \rangle~ \bbe^A
\ee
must vanish. This gives that
\bea
-(p+\lambda) q+(\mu+z^{12}) z^{34}-z^{13} z^{24}+z^{14} z^{23}=0~,
\la{qqq}
\eea
for every $\lambda$ and $\mu$. As a result, one has that
that $q=0$, $z^{34}=0$. A $\times^2SU(2)$ transformation, which leaves $1$ and $e_{12}$ invariant, can  be used to set,  without
loss of generality, $z^{23}=0$. Thus
\be
\epsilon^5= p\,1 +z^{12} e_{12} + z^{13} e_{13}+z^{14} e_{14}+z^{24} e_{24}~.
\ee
Next, computing the 1-form bilinear associated with the Killing vector one finds
\bea
\label{kvc}
\kappa &=&
\langle B (C*(\epsilon^5+\lambda.1+\mu e_{12}))^*, \Gamma_A (\epsilon^5+\lambda.1+\mu e_{12}) \rangle~\bbe^A
\cr
&=&-\sqrt{2} \big(|p+\lambda|^2+|z^{12}+\mu|^2+|z^{13}|^2+|z^{14}|^2+|z^{24}|^2\big)~\bbe^-~.
\eea
Since the dual vector field must be Killing the function   multiplying $\bbe^-$ must be constant for all constants $\lambda, \mu$. This  forces $p$ and $z^{12}$ to be constant as well.
As a result, the 5th Killing spinor can be chosen as
\be
\epsilon^5= z^{13} e_{13}+z^{14} e_{14}+z^{24} e_{24}~.
\ee
Using (\ref{qqq}), one finds that $z^{13} z^{24}=0$. Thus either $z^{13}$ or $z^{24}$ must vanish. In either case, a $\times^2SU(2)$ transformation
can be used, which leaves $1$ and $e_{12}$ invariant, such that  without loss of generality,
\be
\epsilon^5 = \mu e_{13}~.
\ee
Examination of
the component $\kappa_-$ in ({\ref{kvc}}) implies that $|\mu|^2$ is constant. As one can set
$\mu \in \bR$ using an appropriately chosen $\times^2SU(2)$ transformation, one can without loss of generality set
$\mu=1$, and $\epsilon^5=e_{13}$. Therefore a basis of the 6 Killing spinors is $\{1, i\,1, e_{12}, i\, e_{12}, e_{13}, i\, e_{13}  \}$. These spinors
have isotropy group $U(1)\ltimes \bR^8$ in $Spin(9,1)$ or $U(1)$ in $Spin(8)$.

\subsection{Analysis of conditions}
The analysis of the additional conditions imposed by requiring that $\epsilon=e_{13}$ be a Killing spinor proceeds
in exactly the same fashion as for $e_{12}$ in the $N=4$ case. Recall that in the $N=4$ case, having
$e_{12}$ as a Killing spinor forced $h_3=h_4=0$ as a consequence of the $+-$ component of the
Killing spinor and compactness of ${\cal{S}}$. Similarly, requiring that $e_{13}$ be a Killing spinor
implies that $h_2=0$ as well.  Using the same reasoning, and from examination of ({\ref{ysol}}),
we must also have
\bea
Y &=& {1 \over 4} \big(h_1 \bbe^1 + h_{\bar{1}} \bbe^{\bar{1}} \big) \wedge \big(i\bbe^3 \wedge \bbe^{\bar{3}}
+i \bbe^4 \wedge \bbe^{\bar{4}} \big)
\nn
&=&  {1 \over 4} \big(h_1 \bbe^1 + h_{\bar{1}} \bbe^{\bar{1}} \big) \wedge \big(i\bbe^2 \wedge \bbe^{\bar{2}}
+i \bbe^4 \wedge \bbe^{\bar{4}} \big)~.
\eea
It follows that $h_1=0$ as well, so $h=0$. From \cite{iibhor}, it follows that the 5-form must vanish. Moreover, since there
are no Berger manifolds with such holonomy\footnote{This $U(1)$ acts differently on the typical fibre of the tangent bundle than the  $U(1)$  holonomy group of 2-dimensional K\"ahler manifolds.} $U(1)\subset \times^2 SU(2)$, the only
solutions are flat and so the spatial horizon section is $T^8$.
Note also that this is compatible with the existence of a hidden structure with skew-symmetric torsion. In particular, it has been shown in \cite{hps} that 8-dimensional KT manifolds
with holonomy ${\rm hol}(\hat\nabla)\subseteq U(1)\subset \times^2 SU(2)$ have vanishing Riemann curvature.

We remark that the same analysis forces all solutions with $N>6$ to have $h=0$ and $F=0$ as well,
modulo the assumptions (i) and (ii) made in section \ref{ass}.

\vskip 0.5cm
\noindent{\bf Acknowledgements} \vskip 0.1cm
\noindent  GP thanks the Gravitational Physics  Max-Planck Institute at Potsdam and the PH-TH Divison at CERN for hospitality
where parts of this work were done. UG is supported by the Knut and Alice Wallenberg Foundation. JG is supported by the EPSRC grant, EP/F069774/1.
GP is partially supported by the EPSRC grant EP/F069774/1 and the STFC rolling grant ST/G000/395/1.
\vskip 0.5cm



\begin{thebibliography}{99}


\bibitem{ring1}
H. Elvang, R. Emparan, D. Mateos and H. S. Reall,
``A Supersymmetric black ring", Phys. Rev. Lett. {\bf{93}} (2004) 211302 [hep-th/0407065].

\bibitem{gibbons1}
G. W. Gibbons, D. Ida and T. Shiromizu,
``Uniqueness and non-uniqueness of
static black holes in higher dimensions",
Phys. Rev. Lett. {\bf{89}} (2002) 041101 [hep-th/0206049].

\bibitem{rogatko}
M. Rogatko,
``Uniqueness theorem of static degenerate and non-degenerate
charged black holes in higher dimensions", Phys. Rev. {\bf{D67}} (2003) 084025;
hep-th/0302091; ``Classification of static charged black holes in higher dimensions,''
Phys. Rev. {\bf{D73}} (2006), 124027 [hep-th/0606116].

\bibitem{obers1}
R. Emparan, T. Harmark, V. Niarchos and N. Obers,
``World-Volume Effective Theory for Higher-Dimensional Black Holes,"
Phys. Rev. Lett. {\bf{102}} (2009) 191301
 [arXiv:0902.0427 [hep-th]];
 ``Essentials of Blackfold Dynamics;" arXiv:0910.1601 [hep-th].


\bibitem{hh}
  J.~Gutowski and G.~Papadopoulos,
  ``Heterotic Black Horizons,''
  JHEP {\bf 1007}, 011 (2010).
  [arXiv:0912.3472 [hep-th]].

``Heterotic horizons, Monge-Ampere equation and del Pezzo surfaces,''
  JHEP {\bf 1010 } (2010)  084.
  [arXiv:1003.2864 [hep-th]].

\bibitem{kunduri}
  H.~K.~Kunduri, J.~Lucietti,
  ``An infinite class of extremal horizons in higher dimensions,''
[arXiv:1002.4656 [hep-th]].


\bibitem{iibhor}
  U.~Gran, J.~Gutowski and G.~Papadopoulos,
  ``IIB black hole horizons with five-form flux and KT geometry;'' arXiv:1101.1247 [hep-th].

  \bibitem{het}
  U.~Gran, P.~Lohrmann and G.~Papadopoulos,
  ``The Spinorial geometry of supersymmetric heterotic string backgrounds,''
  JHEP {\bf 0602}, 063 (2006).
  [hep-th/0510176].


   U.~Gran, G.~Papadopoulos, D.~Roest and P.~Sloane,
``Geometry of all supersymmetric type I backgrounds,''
  JHEP {\bf 0708}, 074 (2007).
  [hep-th/0703143].

 G.~Papadopoulos,
  ``Heterotic supersymmetric backgrounds with compact holonomy revisited,''
  Class.\ Quant.\ Grav.\  {\bf 27 } (2010)  125008.
  [arXiv:0909.2870 [hep-th]].



  \bibitem{iibn1}
U.~Gran, J.~Gutowski and G.~Papadopoulos,
  ``The spinorial geometry of supersymmetric IIB backgrounds,''
  Class.\ Quant.\ Grav.\  {\bf 22} (2005) 2453
  [arXiv:hep-th/0501177].


  ``The G(2) spinorial geometry of supersymmetric IIB backgrounds,''
  Class.\ Quant.\ Grav.\  {\bf 23} (2006) 143
  [arXiv:hep-th/0505074].


  ``The spinorial geometry of supersymmetric IIB backgrounds,''
  Class.\ Quant.\ Grav.\  {\bf 22}, 2453 (2005)
  [arXiv:hep-th/0501177].

  \bibitem{iib28}
  U.~Gran, J.~Gutowski, G.~Papadopoulos and D.~Roest,
  ``N = 31 is not IIB,''
  JHEP {\bf 0702} (2007) 044
  [arXiv:hep-th/0606049].

  ``IIB solutions with $N>28$ Killing spinors are maximally supersymmetric,''
  JHEP {\bf 0712} (2007) 070
  [arXiv:0710.1829 [hep-th]].

  U.~Gran, J.~Gutowski and G.~Papadopoulos,
  ``Classification of IIB backgrounds with 28 supersymmetries,''
  JHEP {\bf 1001} (2010) 044
  [arXiv:0902.3642 [hep-th]].

   J.~M.~Figueroa-O'Farrill and G.~Papadopoulos,
  ``Maximally supersymmetric solutions of ten-dimensional and eleven-dimensional supergravities,''
  JHEP {\bf 0303 } (2003)  048.
  [hep-th/0211089].

\bibitem{het1}
  U.~Gran, G.~Papadopoulos, D.~Roest,
  ``Supersymmetric heterotic string backgrounds,''
  Phys.\ Lett.\  {\bf B656 } (2007)  119-126.
  [arXiv:0706.4407 [hep-th]].



  \bibitem{hps}
  P.~S.~Howe, G.~Papadopoulos, V.~Stojevic,
  ``Covariantly constant forms on torsionful geometries from world-sheet and spacetime perspectives,''
  JHEP {\bf 1009 } (2010)  100.
  [arXiv:1004.2824 [hep-th]].



\bibitem{west}
J.~H.~Schwarz and P.~C.~West,
``Symmetries And Transformations Of Chiral N=2 D = 10 Supergravity,''
Phys.\ Lett.\ B {\bf 126} (1983) 301.

\bibitem{schwarz}
J.~H.~Schwarz, ``Covariant Field Equations Of Chiral N=2 D = 10
Supergravity,'' Nucl.\ Phys.\ B {\bf 226} (1983) 269.

\bibitem{howe}
P.~S.~Howe and P.~C.~West, ``The Complete N=2, D = 10
Supergravity,'' Nucl.\ Phys.\ B {\bf 238} (1984) 181.



%
%
%
%
%
%
%
%
%
%
%
%


%
%
%
%
%
%
%
%
%
%
%
%
%
%
%
%
%
%
%
%
%
%
%
%
%
%
%
%
%
%
%
%
%
%
%
%
%
%
%
%
%
%
%
%
%
%
%
%
%
%
%
%
%
%
%
%
%
%
%
%
%
%
%
%
%
%
%



\end{thebibliography}
\end{document}